\documentclass[twocolumn,showpacs,preprintnumbers,amsmath,amssymb,
superscriptaddress,nofootinbib]{revtex4}
\usepackage{graphicx}

\newcommand{\eps}{\varepsilon}
\newcommand{\tens}[1]{{\boldsymbol{#1}}}

\newcommand{\Ric}{{\mathbf{Ric}}}

\newcommand{\n}[1]{\label{#1}}
\newcommand{\be}{\begin{equation}}
\newcommand{\ee}{\end{equation}}
\newcommand{\ba}{\begin{eqnarray}}
\newcommand{\ea}{\end{eqnarray}}

\newcommand{\hook}{\raisebox{-0.35ex}{\makebox[0.6em][r]
{\scriptsize $-$}}\hspace{-0.15em}\raisebox{0.25ex}{\makebox[0.4em][l]{\tiny
 $|$}}}

\begin{document}

\title{On the supersymmetric limit of Kerr-NUT-AdS metrics}

\author{David Kubiz\v n\'ak}

\email{dk317@cam.ac.uk}

\affiliation{DAMTP, University of Cambridge, Wilberforce Road, Cambridge CB3 0WA, UK}

\date{February 11, 2009}  

\begin{abstract} 
Generalizing the scaling limit of Martelli and Sparks [hep-th/0505027] into an arbitrary number of spacetime dimensions we 
re-obtain the (most general explicitly known) Einstein--Sasaki spaces constructed by Chen, L\"u, and Pope [hep-th/0604125]. 
We demonstrate that this limit has a well-defined geometrical meaning which links together 
the principal conformal Killing--Yano tensor  of the original Kerr-NUT-(A)dS spacetime, 
the K\"ahler 2-form of the resulting Einstein--K\"ahler base, and the 
Sasakian 1-form of the final Einstein--Sasaki space. 
The obtained Einstein--Sasaki space possesses the tower of Killing--Yano tensors of increasing rank---underlined by the existence of Killing spinors. A similar tower of hidden symmetries is observed in the original (odd-dimensional) Kerr-NUT-(A)dS spacetime.
This rises an interesting question whether also these symmetries can be related to the existence of some `generalized' Killing spinor.   
\end{abstract}

\pacs{04.50.-h, 02.40.-k, 04.50.Gh, 04.20.Jb \hfill  DAMTP-2009-8}

\maketitle

\section{Introduction}
In the last few years, there has been a considerable interest in constructing explicit examples 
of compact Riemannian manifolds admitting Killing spinors. Among them, of primary interest are 
the Einstein--Sasaki spaces which provide supersymmetric backgrounds relevant to the AdS/CFT correspondence \cite{Maldacena:1998}. The most interesting examples recently obtained are the 
infinite families of 5D  
$Y^{p,q}$ \cite{GauntlettEtal:2004, GauntlettEtal:2006} 
and $L^{p,q,r}$ \cite{CveticEtal:2005a, CveticEtal:2005b} 
 spaces and their higher-dimensional generalizations \cite{GauntlettEtal:2004, GauntlettEtal:2006, MartelliSparks:2005, ChenEtal:2005, CveticEtal:2005a, CveticEtal:2005b, ChenEtal:2007, LuEtal:2007, ChenEtal:2006cqg}. 

There are several approaches  to the construction of these Riemannian manifolds.
For example, it turned out that the $Y^{p,q}$ spaces can be obtained by a certain 
scaling limit of the Euclideanised five-dimensional Kerr-(A)dS black hole metrics \cite{HashimotoEtal:2004}. More generally, one can consider an Euclidean analogue of the BPS limit
of the higher-dimensional Kerr-NUT-AdS spacetimes \cite{CveticEtal:2005a, CveticEtal:2005b, ChenEtal:2007, ChenEtal:2006cqg} (see also \cite{CveticEtal:2005d} for the limit in the Lorentzian regime). In this approach, one studies the eigenvalues of the
Bogomol'nyi matrix arising from the supersymmetry algebra, see \cite{CveticEtal:2005c}. The condition for saturating the Bogomol'nyi type bound imposes restrictions on the `charges' of the  Euclideanised Kerr-NUT-AdS metric and suggests a certain limit of the parameters characterizing the solution. The requirement that the resulting metric remains non-trivial determines the necessary additional transformation of coordinates. 
Consequently, the odd-dimensional Kerr-NUT-AdS metrics give rise to the Einstein--Sasaki spaces 
whereas the even-dimensional Kerr-NUT-AdS metrics result in the Ricci-flat K\"ahler manifolds.

A slightly different approach to constructing the Einstein--Sasaki spaces is based on the fact 
that an (odd-dimensional) Einstein--Sasaki space
is in a one-to-one correspondence with a one dimension lower (even-dimensional)  
Einstein--K\"ahler metric (see, e.g., \cite{GibbonsEtal:2003}). 
Namely, let $\tens{g}_{\mbox{\tiny EK}}$ be a $2n$-dimensional Einstein--K\"ahler manifold 
obeying 
\be\label{Ric_EK}
\Ric_{\mbox{\tiny EK}}=(2n+2)\tens{g}_{\mbox{\tiny EK}}\,,
\ee
and 
$\tens{\Omega}$ be its associated K\"ahler 2-form with the potential $\tens{A}$, 
$\tens{\Omega}=\tens{dA}$. Then the $U(1)$ bundle over $\tens{g}_{\mbox{\tiny EK}}\,$, 
\be\label{bundle}
\tens{g}_{\mbox{\tiny ES}}=\tens{g}_{\mbox{\tiny EK}}+(2\tens{A}+\tens{d}\psi_n)^2\,, 
\ee   
is a $(2n+1)$-dimensional Einstein--Sasaki space obeying  
\be\label{Ric_ES}
\Ric_{\mbox{\tiny ES}}=2n \tens{g}_{\mbox{\tiny ES}}\,.
\ee

This property was used by Martelli and Sparks \cite{MartelliSparks:2005} 
to re-construct the $L^{p,q,r}$ spaces, discovered earlier by the BPS limit 
by Cveti\v{c} et al. \cite{CveticEtal:2005a, CveticEtal:2005b}.
In their construction, Martelli and Sparks first 
obtained a family of 4D local toric Einstein--K\"ahler metrics, by taking a certain scaling limit of the Euclideanised form of the Kerr-NUT-(A)dS Carter--Pleba\'nski metric \cite{Carter:1968pl, Carter:1968cmp, Plebanski:1975}, and then constructed the final Einstein--Sasaki space as a $U(1)$ bundle over this metric.

In this paper we want to take a closer look at this approach.
It is now known that the general Kerr-NUT-(A)dS metrics  in $D\ge 3$ spacetime dimensions \cite{ChenEtal:2006cqg} possess a hidden symmetry associated with the principal conformal Killing--Yano (PCKY) tensor \cite{KubiznakFrolov:2007}. We shall show that the scaling limit of Martelli and Sparks 
may be understood as a process in which the original (completely non-degenerate) PCKY tensor becomes a completely degenerate one, with $n=[D/2]$ unit eigenvalues, together with the requirement that the metric remains finite.
In particular, this means that in an even number of dimensions the PCKY tensor degenerates to  
the (covariantly constant) K\"ahler 2-form and the Kerr-NUT-(A)dS spacetime becomes the Einstein--K\"ahler manifold. 
For the $D$ odd the limit results in an Einstein--Sasaki space with a degenerate 
closed conformal Killing--Yano tensor. This tensor is directly related to the corresponding Sasakian 1-form. 

Slightly more generally, one can consider the even(odd)-dimensional canonical metric element admitting the PCKY tensor \cite{HouriEtal:2007, KrtousEtal:2008} and perform the scaling limit to obtain 
the K\"ahler (Sasaki) space---without employing the field equations.
Contrary to the BPS limit, where the primary transformation 
(guided by the supersymmetric algebra) is the transformation of parameters, in 
this approach the resulting manifolds are constructed from the `purely geometrical' point of view.

In Section II, we review some  basic facts about the (even-dimensional) canonical metric admitting the PCKY tensor, provide a motivation for the scaling limit 
leading to the K\"ahler manifold, explicitly perform this limit 
to obtain the Einstein--K\"ahler metric from the Kerr-NUT-(A)dS spacetime, 
and construct the most general explicitly known Einstein--Sasaki space as a $U(1)$ bundle over this metric. An odd-dimensional version of this scaling limit, leading to the same Einstein--Sasaki space, is discussed in the appendix.
The towers of hidden symmetries of the original canonical metric and the obtained Sasaki space are compared in Section III.
Section IV is devoted to conclusions.

\section{Scaling limit of Martelli and Sparks}
\subsection{PCKY tensor and canonical metric}
The PCKY tensor $\tens{h}$ is a non-degenerate closed conformal Killing--Yano 2-form
\cite{Kashiwada:1968, Tachibana:1969}, \cite{KrtousEtal:2007jhep}. 
This means that for all vector fields $\tens{X}$ there exists such a 1-form $\tens{\xi}$ so that\footnote{%
In what follows we use the notations of \cite{FrolovKubiznak:2008, Kubiznak:phd}. 
Operations $\flat$, $\sharp$ correspond to `lowering', `rising' of 
indices of vectors, forms, respectively. $\tens{\delta}$ denotes the
co-derivative. For a $p$-form  $\tens{\alpha}$ one has
$\tens{\delta\alpha}= \epsilon
\tens{*}\tens{d}\tens{*}\tens{\alpha},$ where
$\tens{d}$ denotes the exterior derivative, $\tens{*}$ denotes the Hodge star
operator, and $\epsilon=(-1)^{p(D-p)+p-1}$. The `hook' operator $\hook$ denotes `contraction'. }  
\be\label{PCKY}
\nabla_{X}\tens{h}=\tens{X}^{\flat}\wedge \tens{\xi}\, .
\ee 
The condition of {non-degeneracy} means that in a generic point of the manifold the skew symmetric matrix $h_{ab}$ 
has the maximum possible (matrix) rank and that the eigenvalues of $\tens{h}$ are 
functionally independent in some spacetime domain. In this domain, such eigenvalues may be used as `natural' coordinates (see \cite{KrtousEtal:2008} for more details).
The equation \eqref{PCKY} implies
\be\label{xi}
\tens{dh}=0\,,\quad \tens{\xi}=-\frac{1}{D-1}\tens{\delta h}\,.
\ee
This means that there exists a 1-form, a PCKY potential, so that
\be
\tens{h}=\tens{db}\,.
\ee
The 1-form $\tens{\xi}$ associated with $\tens{h}$ is called {\em primary}.

The most general (off-shell) canonical metric\footnote{%
To stress that a metric does not necessarily satisfy the Einstein equations we call it off-shell.}
 admitting the PCKY tensor was constructed 
in \cite{HouriEtal:2007, KrtousEtal:2008}.
In an even dimension ($D=2n$) the metric and the PCKY potential are 
(we sum over $\mu=1,\dots, n$) 
\ba
\tens{g}_{\mbox{\tiny can}}\!&=&\!\frac{U_\mu \tens{d}x_{\mu}^2}{X_\mu(x_\mu)}
  +\frac{X_\mu(x_\mu)}{U_\mu}\Bigl(\sum_{j=0}^{n-1} A_{\mu}^{(j)}\tens{d}\psi_j\!\Bigr)^{\!2},\ \ \label{KNdS2}\\
\tens{b}\!&=&\!-\frac{1}{2}\sum_{k=0}^{n-1} A^{(k+1)}\tens{d}\psi_k\,,\label{b}
\ea
where  
\begin{gather}
U_{\mu}=\prod_{\nu\ne\mu}(x_{\nu}^2-x_{\mu}^2)\,,\nonumber\\
A^{(k)}_\mu=\!\!\!\!\!\sum_{\substack{\nu_1<\dots<\nu_k\\\nu_i\ne\mu}}\!\!\!\!\!x^2_{\nu_1}\dots x^2_{\nu_k},\quad 
A^{(k)}=\!\!\!\!\!\sum_{\nu_1<\dots<\nu_k}\!\!\!\!\!x^2_{\nu_1}\dots x^2_{\nu_k}\;\label{vztahy}.
\end{gather}
Introducing the basis
\be\label{omega}
\tens{\omega}^{\hat \mu} =\sqrt{\frac{U_\mu}{X_\mu}}\,
\tens{d}x_\mu\,,\   
\tens{\tilde  \omega}^{\hat \mu} = \sqrt{\frac{X_\mu}{U_\mu}}
 \sum_{j=0}^{n-1}A_{\mu}^{(j)}\tens{d}\psi_j\;,
\ee 
the metric and the PCKY tensor take the form
\ba
\tens{g}_{\mbox{\tiny can}}\!\!\!&=&\!\tens{\omega}^{\hat \mu}\tens{\omega}^{\hat \mu}+
\tens{\tilde \omega}^{\hat \mu}\tens{\tilde \omega}^{\hat \mu}\,,\\
\tens{h}\!&=&\!\tens{db}=x_\mu\, \tens{\omega}^{\hat \mu}
\wedge \!\tens{\tilde \omega}^{\hat \mu}\,.\label{PCKY_2n}
\ea
This means that the chosen basis is the Darboux basis for the PCKY tensor and that 
the coordinates $x_\mu$ are natural coordinates associated with its `eigenvalues'
(see \cite{FrolovKubiznak:2008, Kubiznak:phd} for more details).

\subsection{Limit of canonical metric}
We would like to perform a limit in which the canonical metric becomes a 
K\"ahler manifold. Since the K\"ahler 2-form can be considered as a `special' closed conformal Killing--Yano tensor, it is natural to seek the limit in which $\tens{h}\to \tens{\Omega}$. This is achieved 
when all the original (functionally independent) eigenvalues $x_\mu$ of $\tens{h}$ become constant (so that $\nabla_X\tens{\Omega}=0$), equal to one 
(so that $\tens{\Omega}^2=-\tens{I}$).\footnote{%
That such a degenerate closed conformal Killing--Yano tensor will be necessarily associated with the K\"ahler geometry follows directly from the recent explicit construction of the most general metric admitting the closed conformal Killing--Yano 2-form \cite{HouriEtal:2008b, HouriEtal:2008c}.
}
So, we are led to the transformation 
\be \label{x_trans}
x_\mu\to 1-\epsilon x_\mu\,,  
\ee
followed by the limit $\eps\to 0$.
In order to obtain a `reasonable' limit of the PCKY tensor and the metric, 
we perform the following additional transformations ($k=0,\dots, n-1$):
\ba
\epsilon(-2\epsilon)^{k}\Bigl[\,\sum_{l=k}^{n-1}\Bigl(   
\begin{array}{c}
\!\footnotesize{n\!-\!k\!-\!1}\!\\\footnotesize{l\!-\!k\!}
\end{array}\Bigr)\psi_{l}\,\Bigr]\!\!&\to \psi_{k}\,,\ \  \label{psi_trans}\\
X_\mu\to \frac{1}{4}(-2\epsilon)^{n+1}X_\mu\,.&  \label{X_trans}
\ea

It is then straightforward to verify that we obtain\footnote{The transition for the PCKY potential $\tens{b}$ is up to constant terms which have to be dropped before the limit $\epsilon\to 0$ is taken.}
\ba\n{transformation1}
\tens{\omega}^{\hat \mu}\!\!&\to&\!\!\tens{o}^{\hat \mu}\!=\sqrt{\frac{\Delta_\mu}{X_\mu(x_\mu)}}\,\tens{d}x_\mu\,,\nonumber\\
\tens{\tilde \omega}^{\hat \mu}\!\!&\to&\!\!\tens{\tilde  o}^{\hat \mu}\! = \sqrt{\frac{X_\mu(x_\mu)}{\Delta_\mu}}
 \sum_{j=0}^{n-1}\sigma_{\mu}^{(j)}\tens{d}\psi_j\;,\\
\tens{b}\!&\to&\! \tens{A}=\sum_{k=0}^{n-1} \sigma^{(k+1)}\tens{d}\psi_k\,,\nonumber
\ea
where 
\begin{gather}
\Delta_{\mu}=\prod_{\nu\ne\mu}(x_{\nu}-x_{\mu})\,,\nonumber\\
\sigma^{(k)}_\mu=\!\!\!\!\!\sum_{\substack{\nu_1<\dots<\nu_k\\\nu_i\ne\mu}}\!\!\!\!\!x_{\nu_1}\dots x_{\nu_k},\quad 
\sigma^{(k)}=\!\!\!\!\!\sum_{\nu_1<\dots<\nu_k}\!\!\!\!\!x_{\nu_1}\dots x_{\nu_k}\;\label{vztahy2}.
\end{gather}
So, we get
\ba
\tens{g}_{\mbox{\tiny can}}\!&\to&\,\tens{g}_{\mbox{\tiny K}}=\tens{o}^{\hat \mu}\tens{o}^{\hat \mu}+
\tens{\tilde o}^{\hat \mu}\tens{\tilde o}^{\hat \mu}\,,\label{K}\\
\tens{h}\!&\to&\, \tens{\Omega}=\tens{dA}=\tens{o}^{\hat \mu}\wedge \tens{\tilde  o}^{\hat \mu}\,.
\ea
It is easy to check that $\tens{g}_{\mbox{\tiny K}}$ is an (off-shell) K\"ahler metric 
and $\tens{\Omega}$ its K\"ahler 2-form.\footnote{%
The K\"ahler 2-form $\tens{\Omega}$ can be also understood as arising from the scaling limit \eqref{x_trans}--\eqref{X_trans} of an almost K\"ahler 2-form 
$\tens{\tilde \Omega}=\tens{\omega}^{\hat \mu}\wedge \tens{\tilde  \omega}^{\hat \mu}$, 
see \cite{MartelliSparks:2005} for the 4D case, see also Sec. 2.2. in \cite{HamamotoEtal:2007}.
In other words, in this limit the properties of $\tens{\tilde \Omega}$ and $\tens{h}$ `merge' to form $\tens{\Omega}$.
Whereas the first one becomes closed, the latter becomes completely degenerate.
}

\subsection{Limit of Kerr-NUT-(A)dS metrics}
So far, our considerations were `purely geometrical'. By performing a certain scaling limit, we have, starting from the canonical metric element and the PCKY tensor, constructed an (off-shell) K\"ahler metric and the corresponding K\"ahler 2-form. We shall now turn to the particular case of Kerr-NUT-(A)dS spacetime and demonstrate that the above described limit can be realized to obtain the Einstein--K\"ahler metric.  

The canonical metric $\tens{g}_{\mbox{\tiny can}}$ becomes the Kerr-NUT-(A)dS spacetime \cite{ChenEtal:2006cqg, HamamotoEtal:2007}
obeying 
\be\n{EE}
\Ric_{\mbox{\tiny KNS}}=(-1)^n(2n-1)c_{n}\tens{g}_{\mbox{\tiny KNS}}\,,
\ee
for the 
the following choice of metric functions:
\be\n{XX}
X_\mu=\sum_{k=0}^n c_k x_\mu^{2k}-2d_\mu x_\mu\,.
\ee   
One of the $(2n+1)$ constants $c_k$, $d_\mu$, can be 
scaled away, leaving the total number of physical parameters equal to $2n$. These are related to the
mass, NUT charges, rotations, and the cosmological constant, cf. \eqref{EE}. 

In order to perform the scaling limit, \eqref{X_trans}, we re-parametrize $X_\mu$
as 
\be
X_\mu=c_n\prod_{i=1}^{2n}(x_\mu-\alpha_i)-2b_\mu x_\mu\,.
\ee
Here, we understand that only $(n-1)$ of constants $b_\mu$ are nontrivial, we set $b_n=0$, and
that $2n$ roots $\alpha_i$ are subject to the $(n-1)$ constraints,
\be\label{constraint}
\sigma^{(2i-1)}(\alpha)=0\,,\quad i=1,\dots,n-1\,,
\ee  
following from the fact that odd powers in the original expression \eqref{XX}
vanish. 

Now, we can perform the following scaling of (unconstrained) parameters:
\begin{gather}
\alpha_i\to 1-\epsilon\alpha_i\,, \quad i=1,\dots,n+1\,,\nonumber\\
b_\mu\to \frac{1}{4}(-2\epsilon)^{n+1}b_\mu\,.\label{alpha_trans}
\end{gather}
This is accompanied with a transformation of the remaining $\alpha_i$'s which follows 
from constraints \eqref{constraint}.
We denote
\be
\prod_{i=n+2}^{2n}(x_\mu-\alpha_i)\to 2^{n+1}C_n+O(\epsilon)\,,
\ee
where $C_n$ is some constant depending on $n$.\footnote{%
For example, in 4D the constraint \eqref{constraint} constitutes only one equation  
\be
\alpha_1+\alpha_2+\alpha_3+\alpha_4=0\,. \nonumber
\ee
It follows, that the last parameter, $\alpha_4$, transforms according to \cite{MartelliSparks:2005}
\be
\alpha_4\to -3+\epsilon(\alpha_1+\alpha_2+\alpha_3)\,,\nonumber
\ee
and $C_2=1/2$. In $D=6$ Eq. \eqref{constraint} gives two constraints. As a result, one finds $C_3=5/8$. 
}
Performing the transformations \eqref{x_trans} and \eqref{alpha_trans}
we find  
\be\label{X_skoro}
X_\mu\!\to \frac{1}{4}(-2\epsilon)^{n\!+\!1} \Bigl[4c_n C_n\!\prod_{i=1}^{n+1}(x_\mu-\alpha_i)-2b_\mu+O(\epsilon)\Bigr]\,.
\ee
We finally set $c_n=(-1)^n/C_n$ and take the limit $\epsilon\to 0$. The new metric functions take the form
\be\label{X_EK}
X_\mu=-4\prod_{i=1}^{n+1}(\alpha_i-x_\mu)-2b_\mu\,.
\ee
The K\"ahler metric $\tens{g}_{\mbox{\tiny K}}$, \eqref{K}, with these metric functions is an Einstein space obeying \eqref{Ric_EK}. 
The scaling limit \eqref{x_trans}--\eqref{X_trans}, together with \eqref{alpha_trans}, can be considered as a `natural' higher-dimensional generalization of the 4D limit considered by Martelli and Sparks \cite{MartelliSparks:2005}.

Let us make two remarks. First, one can easily perform a different transformation of parameters to obtain the Ricci-flat K\"ahler manifold  instead of the Einstein--K\"ahler one. For example, we set
\ba
c_n&\to&-\epsilon c_n\,,\quad
b_\mu\to \frac{1}{4}(-2\epsilon)^{n+1}b_\mu\,,\nonumber\\
\alpha_{n+1}&\to& 0\,,\ \ \alpha_i\to 1-\epsilon\alpha_i\,, \ \ i=1,\dots,n\,.\qquad \n{cn_trans}
\ea  
Instead of \eqref{X_EK}, we arrive at 
\be\label{flat}
X_\mu=4c_n C_n\prod_{i=1}^{n}(x_\mu-\alpha_i)-2b_\mu\,,
\ee
and the resulting K\"ahler metric \eqref{K} is Ricci-flat. This is no surprise, since the first transformation in \eqref{cn_trans} effectively sets the cosmological constant equal to zero.
Such a limit corresponds to the BPS limit studied in \cite{ChenEtal:2006cqg}.

Second, it is well known that in 4D the canonical metric element, \eqref{KNdS2}, can also describe the charged Kerr-NUT-(A)dS spacetime \cite{Carter:1968pl, Plebanski:1975}. The electromagnetic charges enter one of the metric functions \eqref{XX}, let us say $X_2$, as additional constant terms; $X_2^{(e,g)}=X_2+e^2-g^2$. The electromagnetic potential is
\be
\tens{\varphi}=-\frac{1}{U_2}\!\left[ex_2(\tens{d}\psi_0\!+\!x_1^2\tens{d}\psi_1)\!+\!
gx_1(\tens{d}\psi_0\!+\!x_2^2\tens{d}\psi_1)\right]\,.
\ee
Performing all the previous transformations together with 
\be
e\to -2\epsilon^2 e\,,\quad g\to -2\epsilon^2g\,,
\ee
we obtain the same Einstein--K\"ahler metric as before (the charges in $X_2$ vanish), but with an additional potential  
\be
\tens{\varphi}=-\frac{1}{\Delta_2}\!\left[e(\tens{d}\psi_0\!+\!x_1\tens{d}\psi_1)\!+\!
g(\tens{d}\psi_0\!+\!x_2\tens{d}\psi_1)\right]\,.
\ee 
This potential gives rise to the `electromagnetic field' $\tens{F}=\tens{d\varphi}$ (satisfying both Maxwell equations) with vanishing energy-momentum tensor. Hence, we have a formal solution of the coupled Einstein--Maxwell theory.  
Another such electromagnetic field in any 4D K\"ahler manifold is given by the K\"ahler 2-form $\tens{\Omega}$,
or by the Ricci 2-form $P_{ab}=1/2 R_{ab}^{\ \ cd}\Omega_{cd}$ in the case of the K\"ahler manifold with positive constant curvature \cite{Pope:1982}. 
The 2-form $\tens{F}$ was used by Martelli and Sparks \cite{MartelliSparks:2005} to construct a harmonic, supersymmetry preserving, (2,1)-form on the Calabi--Yau cone over the $L^{p,q,r}$ space (see also \cite{ChenEtal:2007b} and references therein).

\subsection{Einstein--Sasaki spaces}
To summarize our main result, the derived Einstein--K\"ahler metric and the K\"ahler potential are 
\ba\label{Kahler_vysledek}
\tens{g}_{\mbox{\tiny EK}}&=&\frac{\Delta_\mu \tens{d}x_{\mu}^2}{X_\mu}
  +\frac{X_\mu}{\Delta_\mu}\Bigl(\sum_{j=0}^{n-1} \sigma_{\mu}^{(j)}\tens{d}\psi_j\!\Bigr)^{\!2}\,,\nonumber\\
X_\mu&=&-4\prod_{i=1}^{n+1}(\alpha_i-x_\mu)-2b_\mu\,,\\  
\tens{A} &=&\sum_{k=0}^{n-1} \sigma^{(k+1)}\tens{d}\psi_k\,. \nonumber
\ea
The metric is diffeomorphic to the one  obtained in \cite{ChenEtal:2006cqg} by the BPS limit of the odd-dimensional Kerr-NUT-(A)dS spacetime. It is also identical to the Einstein--K\"ahler metric admitting the non-degenerate Hamiltonian 2-form, constructed already in 
\cite{ApostolovEtal:2004}. 

The $U(1)$ bundle over this metric, \eqref{bundle}, is an Einstein--Sasaki space. It is 
diffeomorphic to the most general explicitly known Einstein--Sasaki space constructed in 
\cite{ChenEtal:2006cqg}.
By restricting its parameters one can obtain a complete and nonsingular manifold.

Slightly more generally, one can consider an (off-shell) Sasaki space  
\ba\label{S}
\tens{g}_{\mbox{\tiny S}}&=&\tens{g}_{\mbox{\tiny K}}+(2\tens{A}+\tens{d}\psi_n)^2\,,
\ea
where $\tens{A}$ and $\tens{g}_{\mbox{\tiny K}}$ are given by \eqref{transformation1}
and \eqref{K}.
As shown in the appendix, this space can be obtained by a scaling limit of an odd-dimensional 
canonical metric admitting the PCKY tensor. 
Its Sasakian 1-form $\tens{\eta}=2\tens{A}+\tens{d}\psi_n$, which is  \cite{Semmelmann:2002} a special unit-norm Killing 1-form obeying for all vector fields $\tens{X}$ 
\ba
\nabla_X \tens{\eta}&=&\frac{1}{2}\tens{X}\hook \tens{d\eta}\,,\label{eta1}\\
\nabla_X (\tens{d\eta})&=&-2\tens{X}^\flat \wedge \tens{\eta}\,,\label{eta2}
\ea
is related to the (degenerate) conformal Killing--Yano tensor $\tens{k}=\tens{dA}$ as
\be
\tens{\eta}=-\frac{1}{D-1}\,\tens{\delta k}\,.
\ee
Since $\tens{k}$ can be obtained by the limit of the (odd-dimensional) PCKY tensor $\tens{h}$, we see that the original PCKY tensor gives rise to the Sasakian 1-form $\tens{\eta}$ defining the Sasaki space \eqref{S}.

\section{Hidden symmetries and Killing spinors}
In this section we would like to bring to attention a similar structure of the tower of hidden symmetries in the original (odd-dimensional) Kerr-NUT-(A)dS spacetimes
and the tower of hidden symmetries of the obtained Einstein--Sasaki spaces.\footnote{%
We reserve the phrase `hidden symmetries' for the existence of (conformal) Killing--Yano tensors. 
}
Whereas the latter is underlined by the existence of Killing spinors, it is at the moment unclear whether the first one can be related to some 
generalized notion of such a spinor.
 
\subsection{Hidden symmetries of canonical spacetimes}
The conformal Killing--Yano (CKY) tensor $\tens{k}$ of rank $p$ is a $p$-form which for all vector fields $\tens{X}$ obeys \cite{Kashiwada:1968, Tachibana:1969}
\be\n{CKY_def3}
{\nabla}_{X} \tens{k}={1\over p+1} \tens{X}\hook \tens{d}\tens{k}
-{1\over D-p+1}\tens{X}^{\flat}\wedge
\tens{\delta} \tens{k}\, .
\ee 
That is, it is an antisymmetric object, the covariant derivative of which splits into the 
`exterior' and `divergence' parts. If the first part vanishes the CKY tensor 
is closed. The vanishing of the second term means that we are dealing with  
a Killing--Yano (KY) tensor. Since, under the Hodge duality the exterior part transforms into the divergence part and reversely, the Hodge dual of a closed CKY tensor is a KY tensor and vice versa.

It was demonstrated relatively recently  \cite{KrtousEtal:2007jhep}, that the ($2n+1$)-dimensional canonical spacetime (see the appendix), possesses a tower of CKY tensors of all ranks. This tower can be generated from the corresponding PCKY tensor as follows:
Having a PCKY tensor, or, more generally, a closed CKY 2-form, $\tens{h}=\tens{db}$, one can construct the tower of closed CKY tensors 
($k=1,\dots,n$) 
\be\label{hk}
\tens{h}_k=\tens{h}^{\wedge k}=\underbrace{\tens{h}\wedge \ldots \wedge
\tens{h}}_{\mbox{\tiny{total of $k$ factors}}}\,,
\ee
with the potentials
\be\label{bk}
\tens{b}_k=\tens{b}\wedge (\tens{db})^{\wedge k}\,, \quad \tens{h}_k=\tens{db}_{k-1}\,.
\ee
Their Hodge duals $\tens{f}_k$ are the KY tensors of rank $(D-2k)$,
\be\label{fk}
\tens{f}_k=\tens{*h}_{k}\,.
\ee
This results in the following tower of CKY tensors of increasing rank:
\be\label{tower_can}
\left\{\tens{*}(\tens{db})^{\wedge n}, \tens{db}, \tens{*}(\tens{db})^{\wedge n-1}\!, \tens{(db)}^{\wedge 2}, \dots, \tens{*db},
\tens{(db)}^{\wedge n}\right\}\,.
\ee
In this tower, the first element is a KY 1-form. It is followed by a closed CKY 2-form, 
which is followed by a KY 3-form, and so on; 
CKY tensors of increasing rank are alternatively KY and closed CKY tensors. 
As we shall see below, this is typical for the tower of CKY tensors constructed from a Killing spinor.

\subsection{Hidden symmetries of Sasaki spaces}
The tower of hidden symmetries in Sasaki spaces was
described already in 2002 by Semmelmann \cite{Semmelmann:2002}. 
The structure of this tower is derived from the Sasakian 1-form $\tens{\eta}$. 
Having such a 1-form, 
one can construct the following $(2k+1)$-forms:
\be
\tens{\omega}_k=\tens{\eta}\wedge(\tens{d\eta})^{\wedge k}\,.
\ee
Using \eqref{eta1} and \eqref{eta2}, one can show that these are {\em special} KY tensors obeying for all vector fields $\tens{X}$
\ba
{\nabla}_{X} \tens{\omega}_k&=&{1\over 2k+2} \tens{X}\hook \tens{d}\tens{\omega}_k\, ,\label{omegak1}\\
\nabla_X (\tens{d\omega_k})&=&-2(k+1)\tens{X}^\flat\wedge \tens{\omega}_k\,.\label{omegak2}
\ea
In particular, this implies that $\tens{\omega}_k$ are eigenforms of the Laplace operator corresponding to the eigenvalues $4(k+1)(n-k)$, and that $\tens{\gamma}_k=\tens{d\omega}_{k-1}$ are 
even-rank closed CKY tensors. One has
\be
 \tens{\gamma}_k=\tens{d\omega}_{k-1}=\tens{(d\eta)}^{\wedge k}=\tens{*\omega}_{n-k}\,.
 \ee
The connection to a more general construction \eqref{hk} is through the obvious fact that
$\tens{\gamma}=\tens{d\eta}$ is a closed CKY tensor. 
In the case of the Sasaki space, \eqref{S}, $\tens{\gamma}=\tens{d\eta}=2\tens{dA}=2\tens{k}$, and this CKY tensor can be understood to be `inherited' from the PCKY tensor of the canonical spacetime.

The tower of  hidden symmetries for a Sasaki space is 
\be\label{tower_S}
\left\{\tens{\eta}, \tens{d\eta}, \tens{\eta}\!\wedge \!\tens{d\eta}, \tens{(d\eta)}^{\wedge 2}, \dots, \tens{\eta}\!\wedge\!\tens{(d\eta})^{\wedge n-1}, \tens{(d\eta)}^{\wedge n}\right\}\,. 
\ee
Again, KY tensors alternate closed CKY tensors as the rank increases.
The new feature in this tower is that the KY tensors $\tens{\omega}_k$ are special, obeying \eqref{omegak2}, while, at the same time, they play the role of potentials 
for the even-rank closed CKY tensors $\tens{\gamma}_k$, cf. Eq. \eqref{bk}.

Let us stress that neither the tower of hidden symmetries for the Sasaki space,
\eqref{tower_S}, nor the tower for the canonical metric, \eqref{tower_can}, are subject to  field equations. These hidden symmetries are purely geometrical, irrespective of the 
fact whether the Einstein equations are satisfied or not.

\subsection{Towers of hidden symmetries from Killing spinors}  
Let us now take a closer look at the relationship of hidden symmetries 
and Killing spinors.
A Killing spinor $\tens{\varphi}^{\pm}$ with a Killing number $\lambda$ is a spinor which for all vector fields $\tens{X}$ obeys
\be\label{KS}
D_X \,\tens{\varphi}^{\pm}=\pm i \lambda \tens{X\cdot}\, \tens{\varphi}^{\pm}\,.
\ee  
Here, $D_X$ denotes the covariant derivative on spinors, and $\tens{\cdot}$ stands for the 
Clifford multiplication. 
Eq. \eqref{KS} imposes strict restrictions on the manifold. Namely, in Euclidean signature and 
for $\lambda$ real, the integrability conditions imply that the manifold is a compact Einstein space with positive curvature and special holonomy.

Having Killing spinors $\tens{\varphi}^{\pm}$, one can construct two towers of CKY tensors of increasing rank \cite{Cariglia:2004}. They consist of $p$-forms 
$\tens{\alpha}_p$, $\tens{\beta}_p$, respectively ($p=1,\dots,D-1$), defined on the arbitrary vector fields $\tens{X}_1,\dots, \tens{X}_p$ as
\begin{gather}
\tens{\alpha}_p(\tens{X}_1,\dots,\tens{X}_p)=\bigl<(\tens{X}_1\!\wedge\! \dots \!\wedge\! \tens{X}_p)\tens{\cdot}\tens{\varphi}^+\!,\tens{\varphi}^-\bigr>,\nonumber\\
\tens{\beta}_p(\tens{X}_1,\dots,\tens{X}_p)=
\bigl<(\tens{X}_1\!\wedge\!\dots \!\wedge\! \tens{X}_p)\tens{\cdot}\tens{\varphi}^+\!,\tens{\varphi}^+\bigr>.\quad\label{joj}
\end{gather}
These towers have the following properties:
Inside each tower, tensors of increasing rank are alternatively KY and closed CKY tensors. Namely, even-rank forms $\tens{\alpha}_{2k}$  are KY tensors and odd-rank forms 
$\tens{\alpha}_{2k+1}$ are closed CKY tensors. Moreover, forms $\tens{\alpha}_{2k}$ are potentials for closed CKY tensors $\tens{\alpha}_{2k+1}$; $\tens{\alpha}_{2k+1}\propto \tens{d\alpha}_{2k}$.
The second tower is similar, but now $\tens{\beta}_{2k+1}$ are KY tensors, whereas $\tens{\beta}_{2k+2}\propto \tens{d\beta}_{2k+1}$ are closed CKY tensors.

We see, that the properties of the tower composed of $\tens{\beta}_p$ are precisely what we have observed in the case of Sasaki spaces.
This is not surprising for the Einstein--Sasaki spaces.
It is well known, see, e.g., \cite{GibbonsEtal:2003}, that such spaces admit a pair of conjugate Killing spinors $\tens{\varphi}^{\pm}$, obeying \eqref{KS} with $\lambda =1$. The tower of hidden symmetries is in this case underlined by the existence of these spinors.
However,  while  
Killing spinors of the kind \eqref{KS} exist only in Einstein spaces, we have seen that the tower \eqref{tower_S} is present for any Sasaki space, irrespectively of the field equations. It is a natural 
question to ask whether the construction \eqref{joj} cannot be realized
for all Sasaki spaces. In other words, can one find the `square root' 
of hidden symmetries \eqref{tower_S} in terms of some `generalized Killing spinor'?
And even more interestingly, can this be done in the original canonical spacetimes?
Whereas the explicit examples of metrics presented in this paper cannot give a general answer, they may provide a useful `test-ground' for studying these questions, especially, 
when the transition between these metrics is explicitly in hand.

\section{Conclusions}
The AdS/CFT correspondence relates the properties of Einstein--Sasaki spaces to 
properties of superconformal field theories. 
The construction of such geometries therefore attracts a lot of attention.
The most general explicitly known Einstein--Sasaki spaces were obtained in \cite{ChenEtal:2006cqg} by the BPS limit of Kerr-NUT-(A)dS spacetimes.
In this paper we have suggested an alternative procedure which generalizes the
approach of Martelli and Sparks \cite{MartelliSparks:2005}.
In this procedure, one starts from the canonical spacetime 
admitting the PCKY tensor and performs a scaling limit in which the PCKY tensor becomes completely degenerate, with equal constant eigenvalues. As a result, in an even number of spacetime dimensions the PCKY tensor `degenerates' to the K\"ahler 2-form 
and one obtains a K\"ahler manifold. In an odd dimension the PCKY tensor gives rise to the Sasakian 1-form and the limit of the canonical spacetime results in the Sasaki metric. 

The advantage of this purely geometrical transition is that it allows one to compare the properties of the original spacetimes and the properties of the resulting manifolds.
In particular, we have noticed a similar structure of the towers of hidden symmetries.
Interestingly enough, this similarity is valid off-shell. It might be related 
to the existence of a some kind of `properly generalized' Killing spinor.
Another advantage of this approach is that one gets a better control on the limit of parameters when considering a particular solution of the field equations. One can, for example, 
directly construct a Ricci-flat K\"ahler manifold, or to preserve the 4D electromagnetic field. 
All this indicates that viewing the 
supersymmetric limit of Kerr-NUT-(A)dS metrics from the perspective of the PCKY tensor might be useful and may possibly bring some new insights in the future.

\appendix

\section{Scaling limit in odd dimensions}
In this appendix we perform a scaling limit of the odd-dimensional ($D=2n+1$)
Kerr-NUT-(A)dS spacetimes.
Similar to the main text, we first consider the geometrical limit of the 
canonical metric admitting the PCKY tensor. The metric and the PCKY tensor are \cite{KrtousEtal:2008}
\ba
\tens{g}_{\mbox{\tiny can}}\!\!\!&=&\!\tens{\omega}^{\hat \mu}\tens{\omega}^{\hat \mu}+
\tens{\tilde \omega}^{\hat \mu}\tens{\omega}^{\hat \mu}+
\tens{\omega}^{\hat \epsilon}\tens{\omega}^{\hat \epsilon}\,,\label{can_odd}\\
\tens{h}\!&=&\!\tens{db}=x_\mu\, \tens{\omega}^{\hat \mu}
\wedge \!\tens{\tilde \omega}^{\hat \mu}\,.\label{h_odd}
\ea 
Here, basis forms $\tens{\omega}^{\hat \mu}$, $\tens{\tilde \omega}^{\hat \mu}$ are given by \eqref{omega}, PCKY potential $\tens{b}$ by \eqref{b}, and we have introduced an additional basis 1-form 
\be
\tens{\omega}^{\hat \epsilon} =\frac{S}{\sqrt{A^{(n)}}}
 \sum_{j=0}^{n} A^{(j)}\tens{d}\psi_j\,.
\ee 
In order to `scale away' the constant $S$ we perform the limit in which we first set 
all the eigenvalues $x_\mu$ of $\tens{h}$ equal to some constant $c$. So we perform
($k=0,\dots, n-1$)
\be
\begin{split} 
x_\mu\to c-\epsilon x_\mu\,,\quad X_\mu\to \frac{1}{4}(-2\epsilon)^{n+1}&X_\mu\,,\\
\epsilon\bigl(-\frac{2\epsilon}{c}\bigr)^{k}\Bigl[\,\sum_{l=k}^{n-1}\Bigl(   
\begin{array}{c}
\!\footnotesize{n\!-\!k\!-\!1}\!\\\footnotesize{l\!-\!k\!}
\end{array}\Bigr)\psi_{l}c^{2l}\,\Bigr]\!\!\to &\,\psi_{k}\,,\ \ \\
\sum_{l=0}^n \Bigl(   
\begin{array}{c}
\!\footnotesize{n}\!\\\footnotesize{l}
\end{array}\Bigr)\psi_{l}c^{2l+1}\!\!\to&\,\psi_n\,,
\end{split}
\ee
followed by the limit $\eps\to 0$, cf. Eqs. \eqref{x_trans}--\eqref{X_trans}.
As a result, we obtain 
$\tens{\omega}^{\hat \mu}\to\tens{o}^{\hat \mu},\,
\tens{\tilde \omega}^{\hat \mu}\to\tens{\tilde o}^{\hat \mu},\, 
\tens{b}\to c\tens{A}\,,$  
given by \eqref{transformation1}, 
and in addition, 
\be
\tens{\omega}^{\hat \epsilon}\to\frac{S}{c^{n+1}}\,\tens{\eta}\,,\quad
\tens{\eta}=2\tens{A}+\tens{d}\psi_n\,.
\ee 
Choosing now the constant $c$, so that $c^{n+1}=S$, 
we get the Sasaki space $\tens{g}_{\mbox{\tiny S}}$, \eqref{S},
\be\label{SS}
\tens{g}_{\mbox{\tiny can}}\!\to\,
\tens{g}_{\mbox{\tiny S}}=\tens{g}_{\mbox{\tiny K}}+(2\tens{A}+\tens{d}\psi_n)^2\,.\\
\ee
The PCKY tensor $\tens{h}/c$ results in the degenerate closed CKY tensor
$\tens{k}=\tens{dA}$, and 
the primary 1-form $\tens{\xi}/c$ transforms to the Sasakian 1-form 
$\tens{\eta}$,
\be
-\frac{1}{D-1}\frac{\tens{\delta h}}{c}=\frac{\tens{\xi}}{c}\to \tens{\eta}=-\frac{1}{D-1}\tens{\delta k}\,. 
\ee

In particular, for the Kerr-NUT-(A)dS spacetimes, obeying 
$\Ric_{\mbox{\tiny KNS}}=(-1)^n 2nc_{n}\tens{g}_{\mbox{\tiny KNS}}$,
we have the following metric functions
\cite{ChenEtal:2006cqg, HamamotoEtal:2007}:
\be\n{XX2}
X_\mu\!=\!\sum_{k=1}^n c_k x_\mu^{2k}\!-\!2d_\mu\! -\!\frac{S^2}{x_\mu^2}
=\frac{c_n}{x_\mu^2}\prod_{i=1}^{n+1}(x_\mu^2\!-\!\alpha_i)\!-\!2b_\mu\,,
\ee
where, again, we set $b_n=0$.
We transform
\be\label{bmu_trans2}
\alpha_i\to c^2-2\epsilon c\alpha_i\,,\quad b_\mu\to \frac{1}{4}(-2\epsilon)^{n+1}b_\mu\,.
\ee
Then we have 
\be\label{X_skoro2}
X_\mu\!\to \frac{1}{4}(-2\epsilon)^{n\!+\!1}\! \Bigl[4c_n c^{n\!-\!1}\prod_{i=1}^{n+1}(x_\mu-\alpha_i)-2b_\mu+O(\epsilon)\Bigr]\,.
\ee
Choosing $c_n=(-1)^n/c^{n-1}$, and in the limit $\epsilon\to 0$,
we find the metric functions \eqref{X_EK}  
and the Sasaki space \eqref{SS} becomes the Einstein--Sasaki space 
obtained in Section II as a $U(1)$ bundle over the Einstein--K\"ahler metric. 

The Sasaki space \eqref{SS} does not allow the Ricci-flat solution (see, e.g., 
\cite{GibbonsEtal:2003}). However, one can perform a slightly different limit of the odd-dimensional canonical metric \eqref{can_odd},  to obtain a 
`K\"ahler string',
\be\label{SSS}
\tens{g}_{\mbox{\tiny can}}\!\to\,
\tens{g}_{\mbox{\tiny string}}=\tens{g}_{\mbox{\tiny K}}+ \tens{d}\tau^2\,.
\ee
The string is Ricci-flat when $\tens{g}_{\mbox{\tiny K}}$ is, that is, when the metric functions take the form \eqref{flat}.   
Metrics \eqref{SS} and \eqref{SSS} correspond precisely to the two possibilities available for the odd-dimensional metric admitting a completely degenerate closed CKY tensor  
\cite{HouriEtal:2008b, HouriEtal:2008c}. In terminology of these papers, the first one corresponds to the `special' type whereas the second one to the `general' type.


\section*{Acknowledgments}
The author is grateful to G.W. Gibbons and H.K. Kunduri for valuable discussions and reading the manuscript, Y. Yasui for email correspondence, and 
the Herchel Smith Postdoctoral Research Fellowship at the University of Cambridge for financial support. 


\end{document}